\documentclass[twocolumn,showpacs,preprintnumbers,amsmath,amssymb]{revtex4}
\usepackage{graphicx}% Include figure files
\usepackage{dcolumn}% Align table columns on decimal point
\usepackage{bm}% bold math

\begin{document}

\preprint{APS}

\title{Noise induced phase transition in the $S$-state block voter model}

\author{J. M. de Ara\'{u}jo$^1$, C. I. N. Sampaio Filho$^2\footnote{Correspondence to: cesar@fisica.ufc.br}$, F. G. B. Moreira$^1$}

\affiliation{$^1$Departamento de F\'{i}sica Te\'{o}rica e  Experimental, Universidade Federal do Rio Grande do Norte, 59072-970, Natal-RN, Brazil\\
$^2$Departamento de F\'{i}sica, Universidade Federal do Cear\'a, 60451-970, Fortaleza-CE, Brazil}

\date{\today}

\begin{abstract}

We use Monte Carlo simulations and finite-size scaling theory to investigate the phase transition and critical behavior of the $S$-state block voter model on square lattices.  It is shown that the system exhibits an order-disorder phase transition at a given value of the noise parameter,  which changes from a continuous transition for $S\le4$ to a discontinuous transition for $S\ge5$. Moreover, for the cases of continuous transition, the calculated critical exponents indicate that the present studied nonequilibrium model system is in the same universality class of its counterpart equilibrium two-dimensional S-state Potts model. We also provide a first estimation of the long-range exponents governing the dependence on the range of interaction of the magnetization, the susceptibility, and the derivative of Binder's cumulant.
\end{abstract}

\pacs{64.60.ah, 64.60.al, 05.50.+q, 89.75.Da}% PACS, the Physics and Astronomy   
                                          
\maketitle

\section{Introduction}

Statistical physics has been used to study social dynamics when we are looking for the simplest and most important properties exhibited by a given system. Indeed, since the qualitative properties of large-scale phenomena do not depend on the microscopic details of the process, only higher level features, such as symmetries, topologies, or conservation laws, are relevant for the global behavior \cite{castellano2009}. The identification of influential spreaders \cite{kitsakNatPhys2010,satorrasRevMod2015}, the creation process of social networks \cite{gallosPhysRevX2012,sampaioSREP2015,clausetPhysRevX2016}, how the opinions and extreme opinions are formed \cite{ramosSREP2015,sibonaPRE2015,calvaoJStat2016}, and how the emergence of consensus is obtained \cite{rednerPRL2003,stanleyPRL2009,graciaPRE2012,janosPRL2013,stanleyPRE2014} are some subjects where the statistical physics finds a plethora of applications.

To address the question about the emergence of a majority-state when multiple states are possible, we consider the collective behavior of the $S$-state block voter model (BVM) \cite{sampaio2011}, which introduces long-range interactions in the system. The BVM is defined by an outflow dynamics where a central set of $N_{PCS}$ spins, denoted by persuasive cluster spins (PCS), tries to influence the opinion of their neighboring counterparts. However, a given spin being influenced offers a persuasion resistance measured by the noise parameter $q$, the probability that a spin adopts a state contrary of the majority of the spins inside the PCS. Precisely, in this work we perform numerical simulations on two-dimensional square lattices of the  $S$-state BVM model, for $S=3,4,5,6,10$, and $20$,  in the $N_{PCS}\times q$ parameter space, where $q$ and $N_{PCS}$ may be regarded as the social temperature and social pressure of the system, respectively. 

After Ref. \cite{castellano2009}, consensus is defined by a configuration in which all agents share the same state. Therefore, due to the presence of the noise $q$, the consensus state is never reached for the $S$-state BVM, except for $q=0$. Indeed, only polarizations and fragmentations are observed for $q\neq0$. Polarization happens when many states are possible but only two of them survive in the population. And fragmentation indicates a configuration with more than two surviving states. 

The BVM does not satisfy the condition of detailed balance, and therefore the zeroth law of thermodynamics is not satisfied. This feature is shared with other studied irreversible models \cite{dickman1999,oliveira1991,oliveira1992,lima2005,marsili2005,sznajd2006,mendes2008,castellano2015,sampaioPRE2016,chen2017}. The parameter $N_{PCS}$ defines the range of interactions and we consider the scenario of medium-range interactions \cite{sampaio2011, sampaio2013}. For the case where only two states are possible, the BVM shows a continuous  order-disorder phase transition with critical exponents described by Ising universality class. Moreover, the long-range exponents that govern the decay of the critical amplitudes of the magnetization, the susceptibility, and the derivative of Binder's cumulant, with the range of interaction, were also calculated \cite{luijtenPRL1996,luijtenPRB1997,blote2002,lubeckPRL2003,sampaio2013}. 
Here, we extend the study of  \cite{sampaio2013} to determine the phase diagram and critical behavior of the block voter model with more than two states.

The remainder of the paper is organized as follows. In Section II we describe the main features of the $S$-state block voter model dynamics used to determine the time evolution of the spin variables associated to each vertex defined on regular lattices. In Section III the results of our simulations are presented and the finite-size scaling analysis is used to investigate the critical properties of the model. We conclude in Section IV. 

\section{The $S$-state block voter dynamics}

The $S$-state block voter model is defined by a set of $N$ spins, where the spin variable $\sigma_{i}$ is associated with the  i-th vertex of a regular square lattice of linear size $L = \sqrt{N}$. Each spin can have $S$ values $\sigma_{i} =1, 2, 3, ...,S$, corresponding to the $S$ possible opinions in a ballot. 
Starting  from a given spin configuration with periodic boundary conditions in both directions,  
the system evolves in time according to the following rules. 
Firstly, a square block consisting of $N_{PCS}$ persuasive cluster spins  $(PCS)$ is randomly chosen and  the majority state of this block is determined. 
Next, a randomly chosen spin located at the adjacency of $PCS$ has its state updated: With probability $(1-q)$ the new state of the adjacent spin agrees with the $PCS$ majority state and it disagrees with probability $q$. 
Notice that, both the noise $q$ and the size $N_{PCS}$ of the persuasive block spin are fixed in time. In this two-parameter model, the parameter $q$ can be viewed as a social temperature defining the resistance to persuasion of the $PCS$, whereas $N_{PCS}$ determines the power of persuasion and it can be thought as a measure of social pressure. 
Moreover, different tie configurations must be considered. In the case of a tie among the $S$ possible states, each state is chosen with equal probability $1/S$. 
In the case of a tie between $M$ majority states, $M=2, ..., (S-1)$, the adjacent spin assumes each one of these states with equal probability $(1-q)/M$, and each one of the other $(S-M)$ states with probability $q/(S-M)$.
The above rules were first used to study the three-state majority-vote model on random graphs  \cite{bradyJSTAT2010}.

To account for the phase diagram and critical behavior of the model in the $N_{PCS} \times q$ parameter space, we consider the magnetization $M_{L}$, the susceptibility $\chi_{L}$, and the Binder fourth-order cumulant $U_{L}$, which are defined by

\begin{equation}
 M_{L}(q) =  \left< \left<  m  \right>_{time}\right>_{sample}, 
 \label{eq01}
\end{equation}

\begin{equation}
 \chi_{L}(q) = N \left[\left< \left< m^{2} \right>_{time}  - \left< m \right>_{time}^{2} \right>_{sample}\right],
 \label{eq02}
\end{equation}

\begin{equation}
 U_{L}(q) = 1 - \left< \frac{\left< m^{4} \right>_{time}}{3\left< m^2 \right>_{time}^{2}} \right>_{sample},
 \label{eq03}
\end{equation}
where the symbols $<\cdots>_{time}$ and $<\cdots>_{sample}$, respectively, denote time averages taken in the stationary state and configurational averages taken over several samples, and $N$ is the number of spins. In the above equations, $m$ is defined in analogy to the magnetization in the $S$-state Potts model as the modulus of the magnetization vector, such that $m = \left(m_{1}^{2} + m_{2}^{2} + ... +m_{S}^{2} \right)^{1/2}$, whose components are given by

\begin{equation}
m_{\alpha} = \sqrt{\frac{S}{(S-1)}}\left[\frac{1}{N}\sum_{i}\delta(\alpha,\sigma_{i}) - \frac{1}{S} \right],
\label{eq04}
\end{equation}
where the summation is over all sites of the lattice, $\delta (\alpha,\sigma_{i})$ is the Kronecker delta function, and the factor $\sqrt{S/(S-1)}$ is introduced in order to normalize the magnetization vector.

For a system with a given value of $S$, we have performed Monte Carlo simulations on regular square lattices of sizes $L = 100,140,180, 220,$ and $280$. In all cases the size of the persuasive cluster spin varies in the range $4 \leq N_{PCS} \leq 100$. Time is measured in Monte Carlo step (MCS), and considering the case of asynchronous update, one MCS corresponds to $N$ attempts of changing the states of the spins.  We wait $10^5$ MCS for the system to reach the steady state and the time averages are calculated based on the next $4\times 10^{5}$ MCS. For all set of parameters $\left( q,N_{PCS}\right)$, at least $100$ independent samples are considered in the calculation of the configurational averages. Moreover, the simulations were performed using different initial spin configurations. 

\section{Results and Discussions}

\begin{figure}[t]
\includegraphics*[width=\columnwidth]{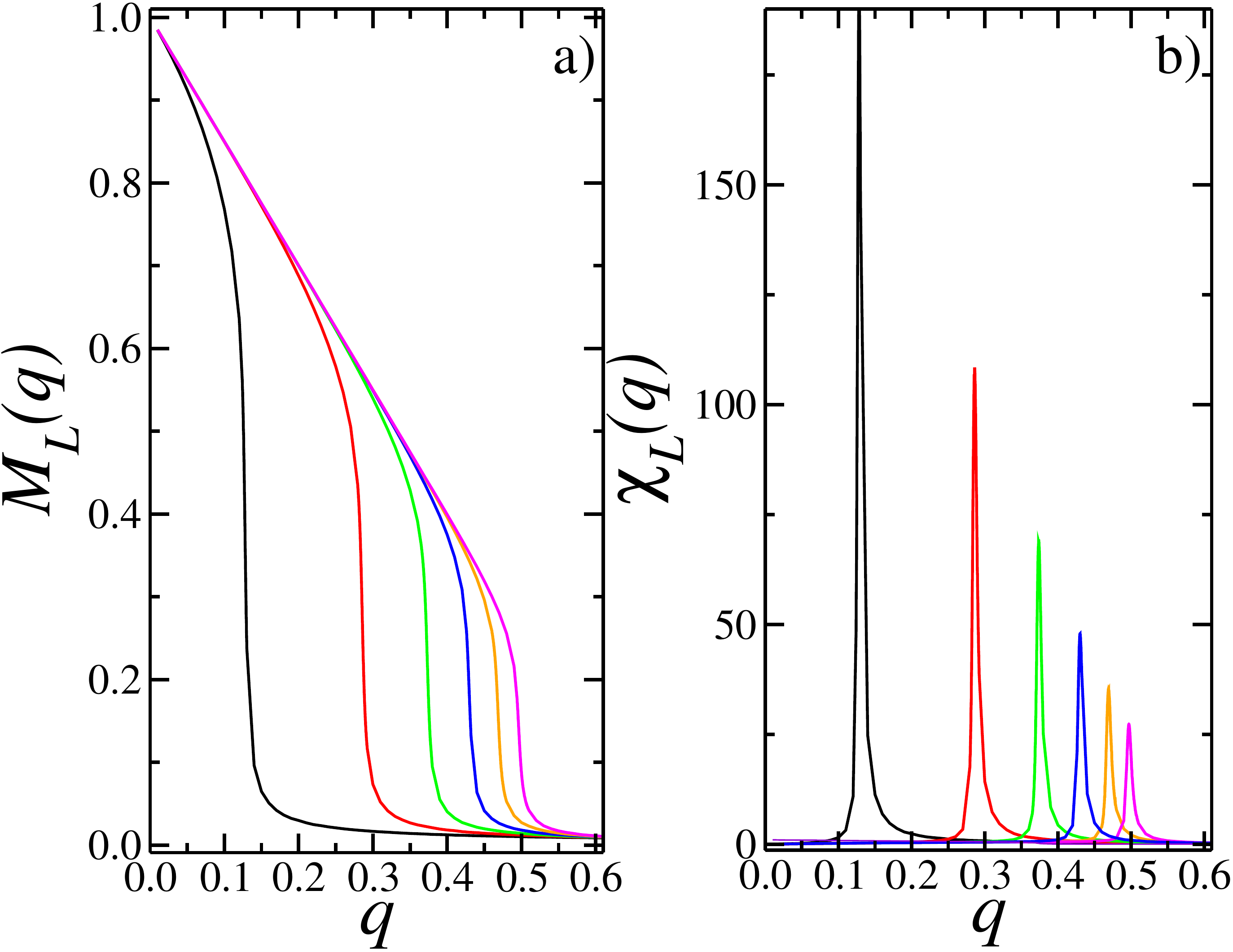}
\caption{Magnetization ($a$) and susceptibility ($b$) as functions of the noise parameter $q$ for $L=100$, and values of $N_{PCS} = 4,9,16,25,36,$ and $49$ (from left to right). Both results suggest that the three-state block voter model undergoes a continuous order-disorder phase transition at a specific value $q_{c}$. Moreover, the peak of the susceptibility becomes more reduced as $N_{PCS}$ increases, implying in the reduction of the critical fluctuations.}
\label{fig01}
\end{figure} 

In Fig.~\ref{fig01} we plot the order parameter $M_{L}$ and the susceptibility $\chi_{L}$ as functions of the noise parameter, for the system with $S=3$. The data were obtained from Monte Carlo simulations on square lattices of size $L=100$ with periodic boundary condition, considering values of $N_{PCS} = 4,9,16,25,36,$ and $49$ (from left to right). We see that the system undergoes an order-disorder phase transition at a value of the noise parameter $q_{c}(N_{PCS})$, which is an increasing function of the size of the persuasive cluster.
Moreover, Fig.~\ref{fig01}(b) shows that the critical amplitudes are reduced as the value of  $N_{PCS}$ increases. In the thermodynamic limit ($N \rightarrow \infty$), we expect the system to show nonzero magnetization only below the critical noise $q_{c}(N_{PCS})$. For finite systems, however, the critical parameter $q_{c}(L)$ for a given $N_{PCS}$ is estimated as the value of $q$ where the corresponding  curve of the susceptibility $\chi_{L}$ in Fig.~\ref{fig01}(b) has a maximum.

\begin{figure}[t]
\includegraphics*[width=\columnwidth]{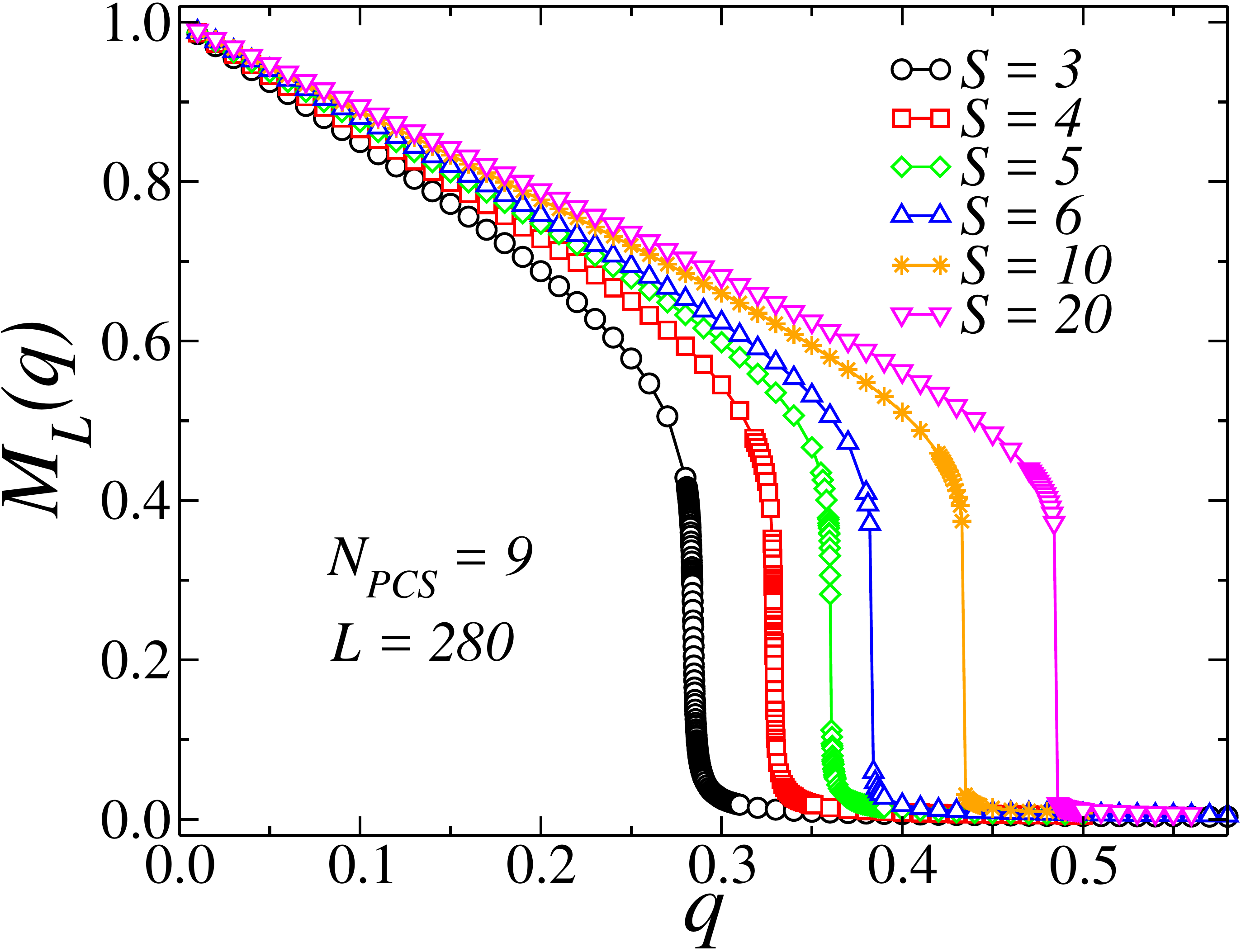}
\caption{The order parameter as a function of noise, for different values of $S$. The data are for $L=280$ and  $N_{PCS} =9$ fixed. A continuous transition in the system ordering is clearly observed for $S=3$ and $S=4$, while for $S\geq 5$ the results suggest a discontinuous transition.}
\label{fig02}
\end{figure}

Fig.~\ref{fig02}  shows the magnetization for different values of $S$ with $L=280$ and  $N_{PCS} =9$ fixed. The simulation results for $S=3$ and $S=4$ clearly indicate that the order-disorder transition is continuous, a feature also observed for the two-state block voter model \cite{sampaio2011}. On the other hand, the results for $M_{L}$ show that the transition changes to discontinuous when $S\geq 5$.

\begin{figure}[t]
\includegraphics*[width=\columnwidth]{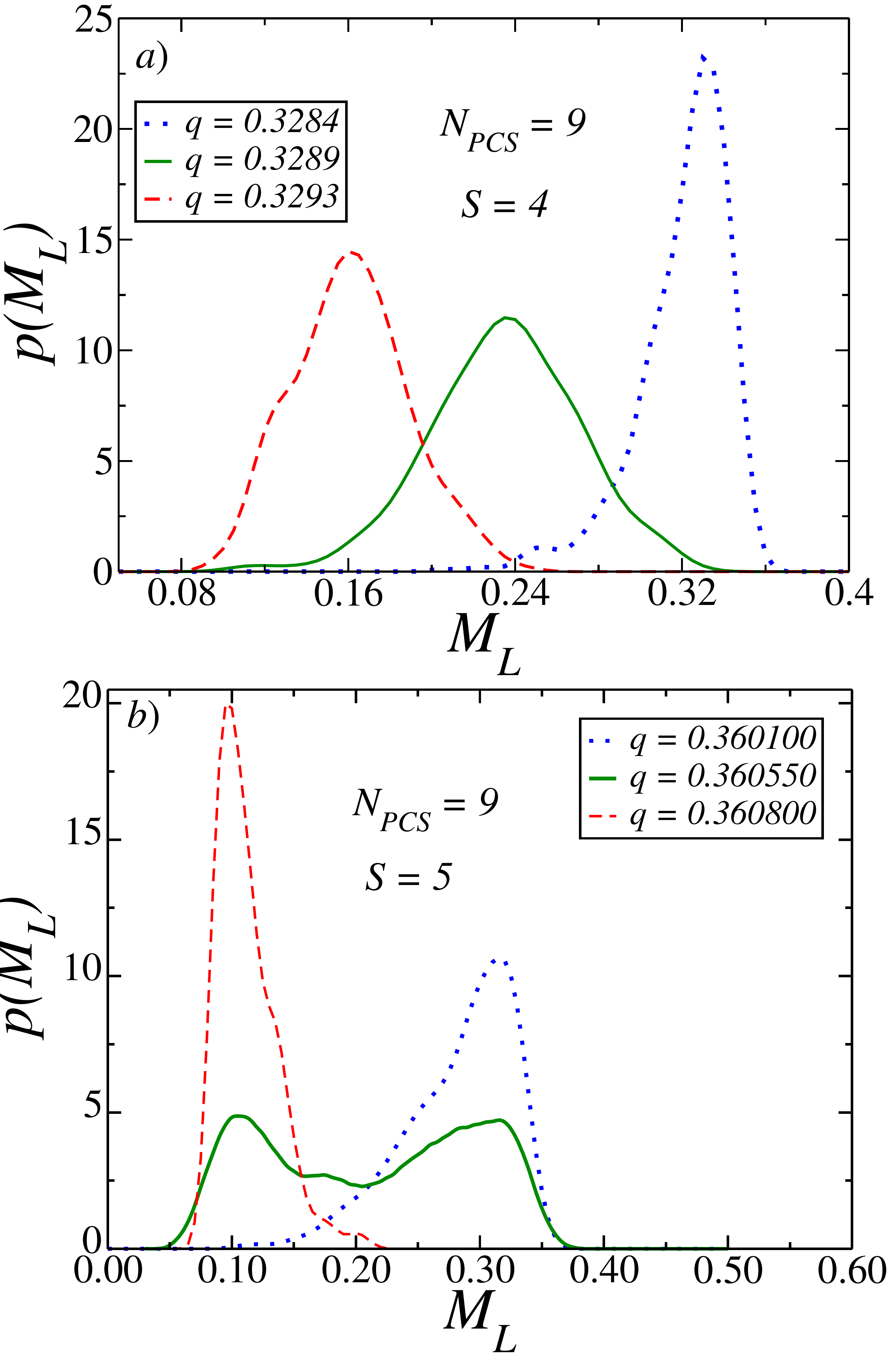}
\caption{Probability density function of the magnetization of S-state block voter model for $N_{PCS}=9$ and $L=280$. $(a)$ Histograms for $S=4$: each curve for $p(M_{L})$ presents a single maximum indicating that transition is continuous. $(b)$ Histograms for $S=5$: the curve in $q=q_c(L) = 0.3605$ (dotted line) exhibits two maxima corresponding to coexisting solutions, while the curves for $q<q_c(L)$ (continuous line) and $q>q_c(L)$ (dashed line) show just one maximum corresponding to the ordered and disordered solutions, respectively. These properties indicate that transition is discontinuous.}
\label{fig03} 
\end{figure}

The nature of the phase transition is better illustrated by the probability density function ($p(M_{L})$) of the order parameter shown in Fig.~\ref{fig03}, obtained from simulations on $700$ samples with $L=280$ and $N_{PCS}=9$. In Fig.~\ref{fig03}(a), the histograms for $S=4$ and for three values of the noise parameter $q$ within the critical region, show that each curve for $p(M_{L})$ presents a single maximum indicating that the transition is continuous. On the contrary, in Fig.~\ref{fig03}(b), the histogram for $S=5$ and $q=q_c(L)=0.3605$ (dotted line) exhibits two maxima corresponding to two coexisting solutions for the order parameter. This must be compared with the curves for $q<q_c(L)$ (continuous line) and $q>q_c(L)$ (dashed line) which present just one maximum corresponding to the ordered and disordered solutions, respectively.

\begin{table}
\vspace{3mm}
\caption{\label{table1} The estimated values of the critical noise $q_{c}$, critical Binder's cumulant $U^{*}$, and critical exponents $\beta/\nu$, $\gamma/\nu$, and $\nu$ for the three-state block voter model on regular square lattice for different values of the parameter $N_{PCS}$. The exponents for the two-dimensional three-state Potts model are $\beta = 1/9$, $\gamma = 13/9$, and $\nu = 5/6$.}
\begin{ruledtabular}
\begin{tabular}{cccccc}
$N_{PCS}$  &$q_{c}$                  &$U^{*}$                 &$\beta/\nu$              &$\gamma/\nu$  &$\nu$ \\
\hline
 $4$       &$0.12630(2)$      &$0.611(2)$        &$0.130(5)$           &$1.70(5)$   &$0.82(4)$  \\
 $9$       &$0.28374(3)$      &$0.609(3)$        &$0.134(2)$           &$1.72(1)$   &$0.83(4)$  \\
 $16$     &$0.37128(2)$      &$0.611(2)$        &$0.143(6)$           &$1.74(1)$   &$0.96(6)$   \\
 $25$     &$0.42760(2)$      &$0.611(2)$        &$0.130(1)$           &$1.72(1)$   &$0.86(2)$    \\
 $36$     &$0.46650(4)$      &$0.610(2)$        &$0.137(2)$           &$1.74(1)$   &$0.81(3)$    \\
 $49$     &$0.49432(3)$     &$0.610(2)$        &$0.136(1)$           &$1.73(1)$   &$0.79(3)$     \\
 $64$     &$0.51555(4)$     &$0.609(2)$        &$0.140(2)$           &$1.74(1)$   &$0.81(3)$     \\
 $81$     &$0.53223(4)$     &$0.608(2)$        &$0.142(2)$           &$1.75(1)$   &$0.81(4)$     \\
 $100$   &$0.54550(2)$    &$0.607(2)$        &$0.141(1)$          &$1.74(1)$   &$0.81(2)$      \\
\end{tabular}
\end{ruledtabular}
\end{table}

\begin{table}
\vspace{3mm}
\caption{\label{table2} The same as Table I for $S=4$. The exponents for the two-dimensional four-state Potts model are $\beta = 1/12$, $\gamma = 7/6$, and $\nu = 2/3$.}
\begin{ruledtabular}
\begin{tabular}{cccccc}
$N_{PCS}$  &$q_{c}$                  &$U^{*}$                 &$\beta/\nu$              &$\gamma/\nu$  &$\nu$ \\
\hline
 $9$       &$0.32852(2)$      &$0.611(2)$        &$0.117(6)$           &$1.77(2)$   &$0.669(  8) $  \\
 $16$     &$0.43018(3)$      &$0.609(2)$        &$0.122(5)$           &$1.74(3)$   &$0.661(6)$   \\
 $25$     &$0.49312(2)$      &$0.609(2)$        &$0.124(4)$           &$1.75(4)$   &$0.667(2)$    \\
 $36$     &$0.53600(3)$      &$0.609(2)$        &$0.126(7)$           &$1.76(1)$   &$0.666(4)$    \\
 $49$     &$0.56670(2)$     &$0.609(1)$        &$0.12(1)$           &$1.75(2)$   &$0.666(6)$     \\
 $64$     &$0.58986(3)$     &$0.609(2)$        &$0.127(7)$           &$1.77(2)$   &$0.668(  8) $     \\
 $81$     &$0.60783(2)$     &$0.610(2)$        &$0.125(2)$           &$1.75(2)$   &$0.664(7)$     \\
 $100$   &$0.62224(2)$    &$0.609(1)$        &$0.12(2)$          &$1.76(2)$   &$0.665(  8) $      \\
\end{tabular}
\end{ruledtabular}
\end{table}

In order to construct the phase diagram for the $S$-state BVM, we have performed the analysis of Binder's cumulant for values of the parameter $N_{PCS}=4,9,16,25,36,49,64,81,100$. 
For a given $S$ and each value of $N_{PCS}$, the critical value $q_{c}(N_{PCS})$ is obtained by calculating the cumulant $U_{L}(q)$, Eq.~(\ref{eq03}), as a function of the noise parameter $q$, considering lattices of sizes $L = 100,140,180, 220,$ and $280$. For sufficiently large system sizes, these curves intercept at a single point $(q_{c},U^{*})$, where $U^{*}=U(q_{c})$. Since the Binder cumulant has zero anomalous dimension \cite{binder1981}, the resulting value of the critical parameter $q_{c}(N_{PCS})$ is independent of $L$. 
Our results for the critical noise $q_c$ and for the critical cumulant $U^{*}$ are presented in Table~I, for $S=3$, and in Table~II, for $S=4$.
As we can notice, there exists a strong dependence between the critical noise and the size of the persuasive cluster, since as $N_{PCS}$ increases the critical noise also increases. 
On the contrary, the value of Binder's cumulant at the intersection $U^{*}$ does not depend (within error bars) on the size of the persuasive cluster.  Considering all set of $N_{PCS}$, we obtain $U^{*}=0.611\pm 0.001$ and  $U^{*}=0.609\pm 0.002$, for $S=3$ and $S=4$, respectively. 
The quoted result for $S=3$ is in agreement with the value $U^{*}=0.61\pm 0.01$ for the equilibrium two-dimensional three-state Potts model and other nonequilibrium three-state models with the same symmetry~\cite{tome2002}. As far as we know there is no previous calculation of $U^{*}$ for $S=4$.

\begin{figure}[t]
\includegraphics*[width=\columnwidth]{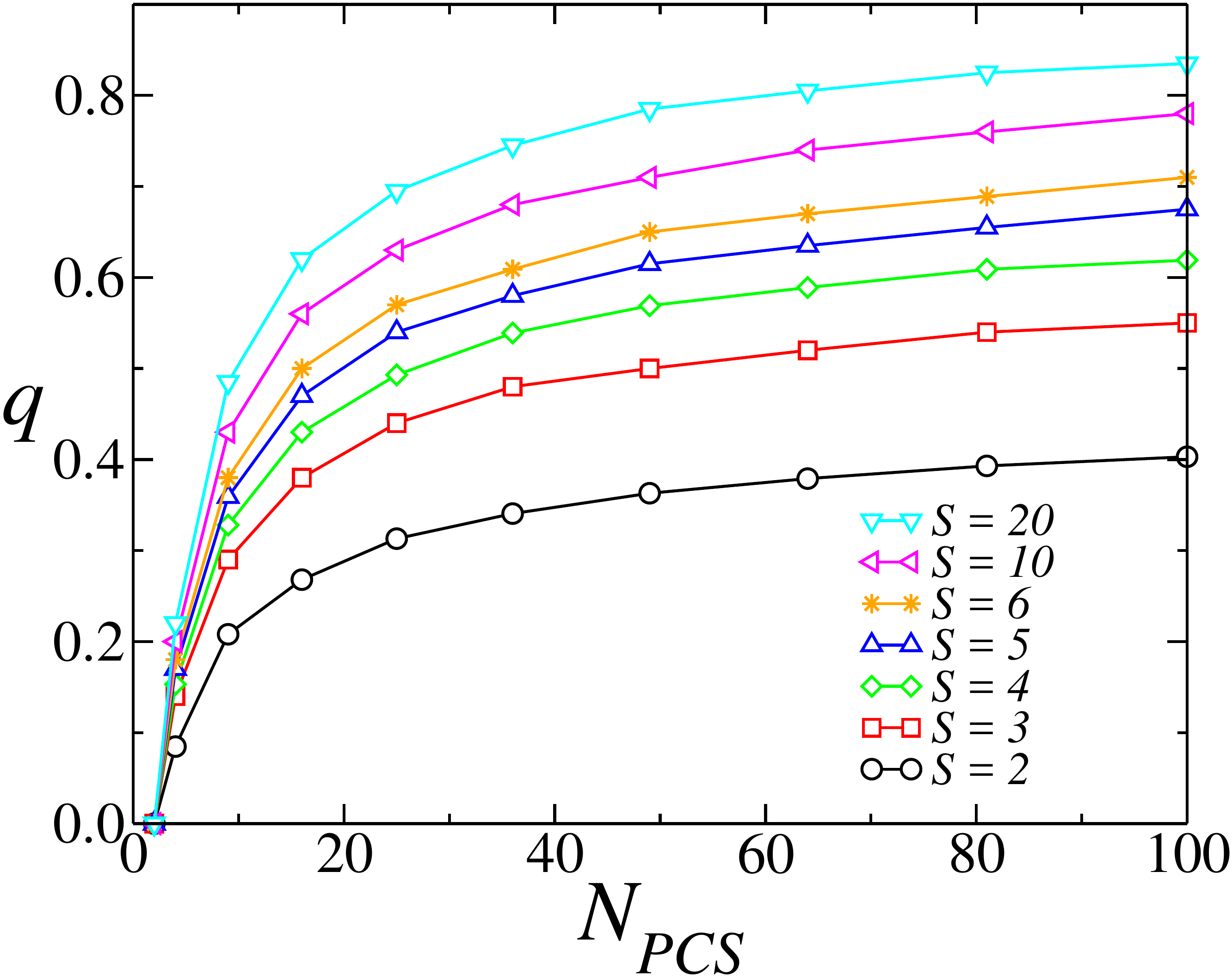}
\caption{The phase diagram of the $S$-state block voter model for $S=2,3,4,5,6$, and $20$ (from bottom to top). Each curve shows that the increasing of the number of spins inside the persuasive cluster favors the ordered phase, which is stable for $q<q_c$. 
The phase diagram for the two-state model (circles) \cite{sampaio2011} is shown for comparison.}
\label{fig04}
\end{figure} 

The dependence of the critical noise $q_{c}$ on the number of spins inside the persuasive cluster
$N_{PCS}$ is given by the phase diagram shown in Fig.~\ref{fig04}. The critical curve, constituted by the critical points obtained from the analysis of Binder's cumulant, separates the ordered phase for $q<q_c$ and disordered phase for  $q>q_c$. For the present studied $S$-state BVM, we see that the critical noise increases monotonically with the size of the persuasive cluster spin. The increase of $q_{c}$ is more pronounced for small values of $N_{PCS}$. This reflects the fact that larger values of $N_{PCS}$ result in larger values of critical noise necessary to destroy the formation of a majority opinion.  It is worth noticing that this only happens for $q<(S-1)/S$, as can be inferred from the condition that the probability $(1-q)$ of a given spin agreeing with the majority state is greater than the probability $q/(S-1)$ of it agreeing with any of the other $(S-1)$ minority states. 
The limit value $q_{c}=(S-1)/S$ corresponds to the mean-field limit $N_{PCS} \approx N$.
Finally we conclude that the increasing of the number of possible states favors the process of formation of a majority opinion, since now one can select one out of $S$ possible states (opinions).

In the following we will use the finite-size scaling theory (FSS) \cite{fisher1972,brezin1981} to study 
the critical behavior of the $S$-state block voter model, for $S=3$ and $S=4$, and to obtain the finite-size dependence of the results of Monte Carlo calculations on finite lattices.
In fact, by performing  the extrapolation of our numerical results to the $N \rightarrow \infty$ limit, we determine the corresponding physical quantities in the thermodynamic limit. This analysis yields good estimates for the  critical exponents and critical parameters, as shown in Table I for $S=3$ and Table II for $S=4$. The calculated exponents $\beta/\nu, \gamma/\nu$, and $\nu$, agree (within error bars) with the quoted values \cite{wu1982} of the critical exponents for the corresponding $S$-state Potts model.
However, the presence of long-ranged interactions described by the parameter $N_{PCS}$ has influence on the nature of both the phase diagram and the critical fluctuations. Therefore, in order to take into account the reduction of the critical amplitudes with increasing range of interactions, we should consider the following ansatz for the scaling equations \cite{binder1993,sampaio2013} 

\begin{equation}
M_{L}(q, N_{PCS}) = N_{PCS}^{-X} L^{-\beta/\nu}\widetilde{M}(\eta),
\label{eq05}
\end{equation}
\begin{equation}
 \chi_{L}(q, N_{PCS}) = N_{PCS}^ {-Y} L^{\gamma/\nu}\widetilde{\chi}(\eta) ,
\label{eq06} 
\end{equation} 
\begin{equation}
u_{L}(q, N_{PCS}) = N_{PCS}^ {-Z} L^{1/\nu}\widetilde{u}(\eta),
\label{eq07}
\end{equation}
where $\widetilde{M}$,  $\widetilde{\chi}$, and $\widetilde{u}$ are universal scaling functions of the scaled variable 
\begin{equation}
\eta = \varepsilon L^{1/\nu} N_{PCS}^{-Z},
\label{eq08}
\end{equation}
$\epsilon = q - q_{c}$ is the distance from the critical noise $q_{c}$ and $\nu$ is the correlation length exponent. The exponents $\beta/\nu$ and $\gamma/\nu$ are associated with the $L$-dependence of the order parameter $M_{L}(q)$ and of the susceptibility $\chi_{L}(q)$, respectively.  Finally, $X$, $Y$, and $Z$ are, respectively, nonnegative exponents governing the dependence on $N_{PCS}$ of the critical amplitudes of the magnetization, the susceptibility \cite{binder1993}, and the derivative of Binder's cumulant $u_{L}(q, N_{PCS}) = \frac{dU}{dq}$. 

\begin{figure}[t]
\includegraphics*[width=\columnwidth]{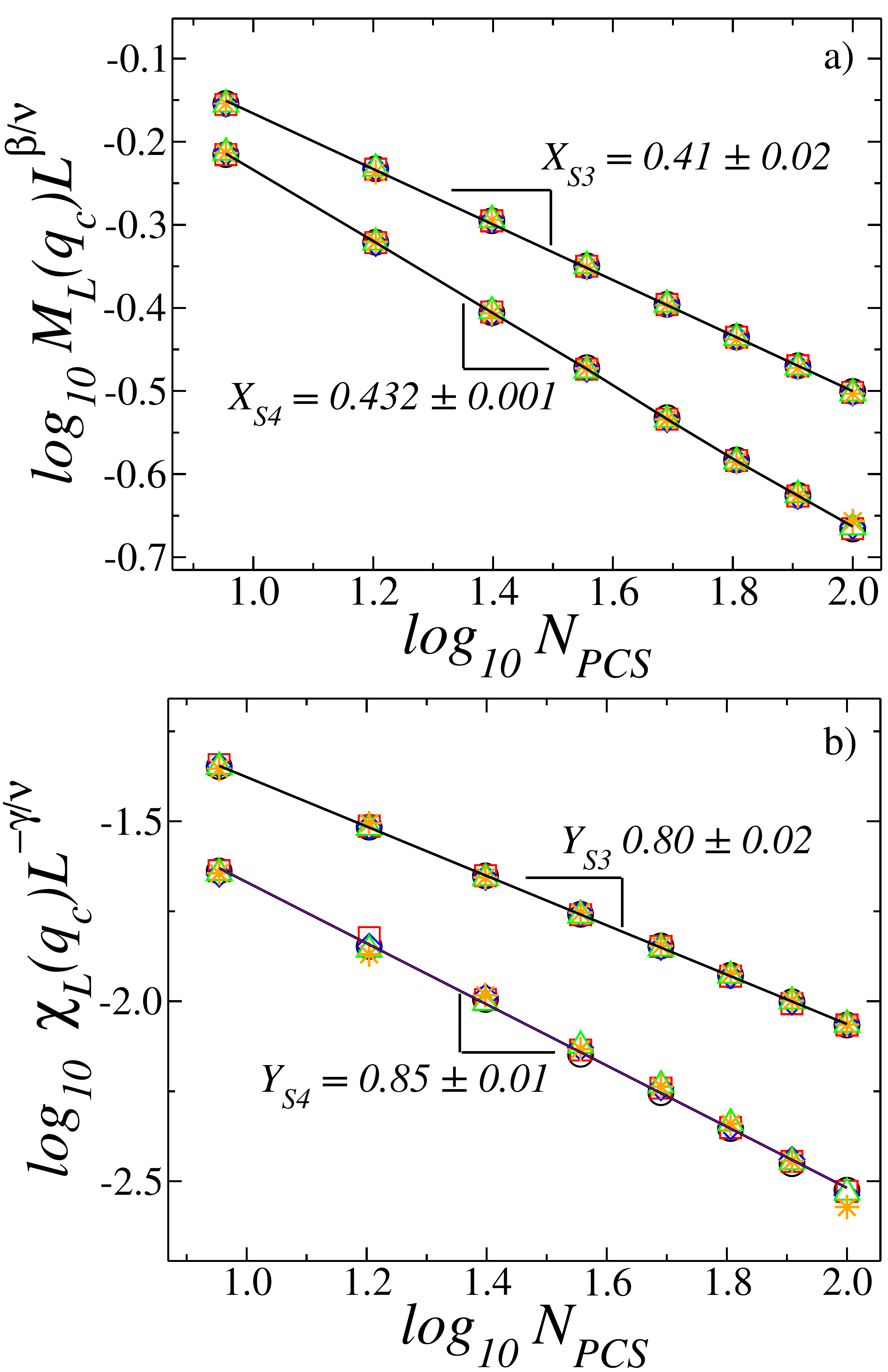}
\caption{The dependence on $N_{PCS}$  of (a) the scaled magnetization and (b) the scaled susceptibility, measured at $q_{c}$. The straight lines represent the scaling relations $M_{L} \sim N_{PCS}^{-X}$ and $\chi_{L} \sim N_{PCS}^{-Y}$, whose slopes yield $X = 0.41\pm 0.02$ and $Y = 0.80\pm 0.02$, for $S=3$, and $X = 0.432\pm 0.001$ and $Y = 0.85\pm 0.01$, for $S=4$.}
\label{fig05}
\end{figure} 

{Figure~\ref{fig05} shows in a $log$-$log$ plot the dependence on the size of the persuasive cluster  spin of the scaled magnetization $L^{\beta/\nu}M_{L}$ (Fig.~\ref{fig05}(a)) and of the scaled susceptibility  $L^{-\gamma/\nu} \chi_{L}$ (Fig.~\ref{fig05}(b)), both calculated at the critical noise $q_{c}$. 
For $S=3$ and $4$, respectively, we use the following set of exponents: $\beta = 1/9$, $\gamma=13/9$, $\nu = 5/6$, and $\beta = 1/12$, $\gamma = 7/6$, $\nu = 2/3$, 
which are the critical exponents for the corresponding two-dimensional $S$-state Potts model~\cite{wu1982}.  The symbols for each value of $N_{PCS}$ represent the values of $M_{L}$ and $\chi_{L}$  for $L= 100, 140, 180, 220$ and $280$. 
The straight lines support the scaling relations $M_{L} \sim N_{PCS}^{-X}$ and $\chi_{L} \sim N_{PCS}^{-Y}$, and their slopes yield  $X=0.41\pm 0.02$ and $Y= 0.80\pm 0.02$, for $S=3$, and $X=0.432\pm 0.001$ and $Y= 0.85\pm 0.01$, for $S=4$.

The exponent $Z$ can be calculated from the requirement that, in the critical region, the scaled variable (Eq.~\ref{eq08}) $|\eta|\sim 1$. This is shown in the $log$-$log$ plot of Fig.~\ref{fig06}, where the slopes of the straight lines obtained from a linear fit to the data yields the exponents $Z=0.05\pm 0.03$ and $Z=0.12\pm 0.03$, respectively, for $S=3$ and $S=4$.

\begin{figure}[t]
\includegraphics*[width=\columnwidth]{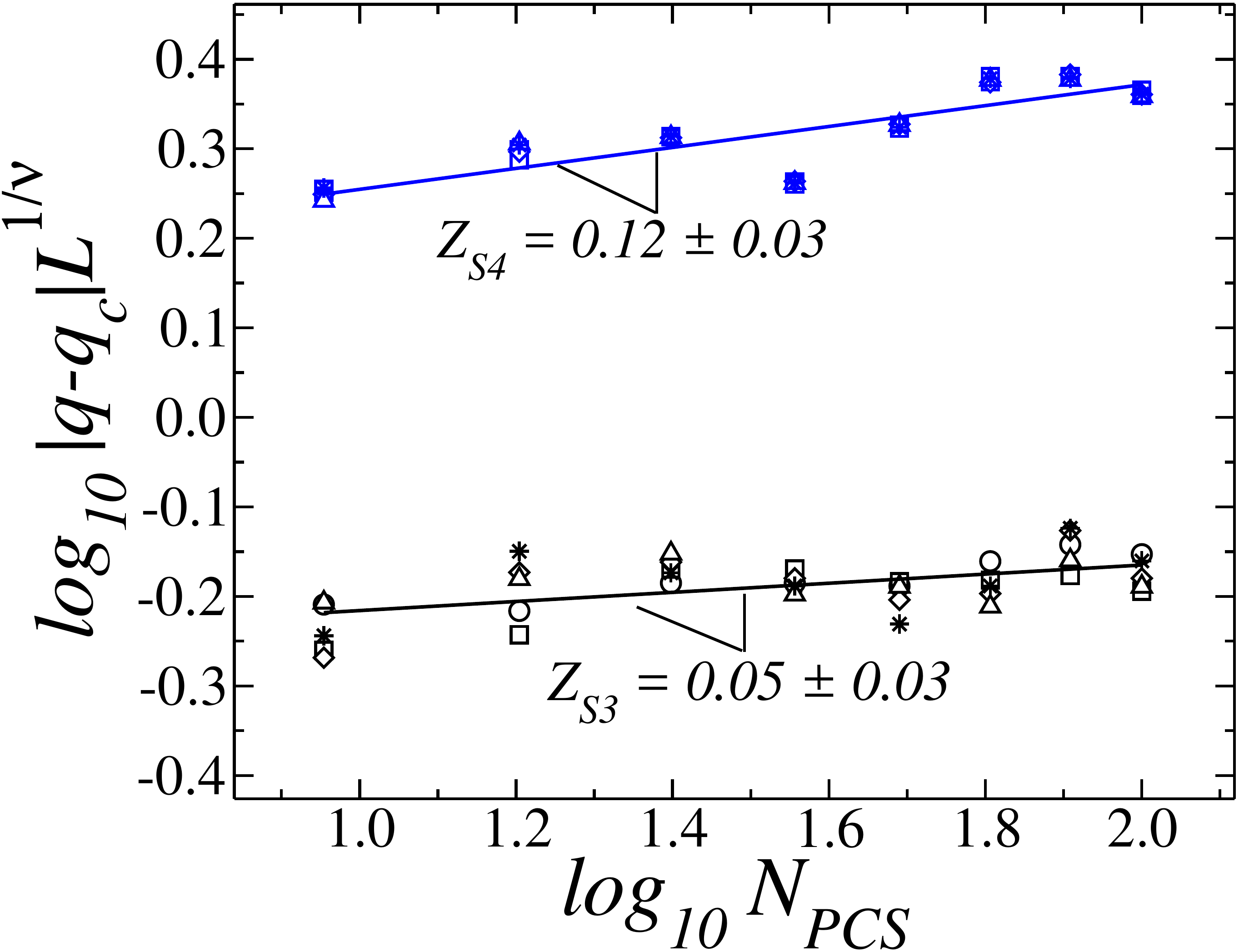}
\caption{Logarithmic plot showing the calculation of the exponent $Z$ for $S=3$ and $S=4$. For each value of $N_{PCS}$ we have five lattice sizes $L = 100$ (circles), $140$ (squares), $180$ (diamonds), $220$ (triangles), and $280$ (stars). The straight lines represent the scaling relation $\epsilon L^{1/\nu} \sim N_{PCS}^{Z}$, whose slopes yields $Z = 0.05 \pm 0.03$ and $Z=0.12\pm 0.03$, respectively, for $S=3$ and $S=4$.} 
\label{fig06}
\end{figure}

\section{Conclusions}
We performed  Monte Carlo simulations and finite-size scaling analysis to obtain the phase diagram and critical exponents of the $S$-state BVM on two-dimensional square lattices. The resulting phase diagram for a given value of $S$, indicates that the increasing of the size of the persuasive cluster favors the ordered phase, that is, the region 
where it is possible to determine a majority state (opinion). 
We found that the order-disorder phase transitions are continuous for $S\leq4$ and discontinuous for $S\geq5$, in agreement with it was observed for the corresponding equilibrium $S$-state Potts model. Our estimates for the critical exponents $\beta, \gamma$, and $\nu$, 
calculated along the lines of continuous phase transitions in the $q \times N_{PCS}$ parameter space, support the conclusion that the $S$-state BVM is in the same universality class of the equilibrium $S$-state Potts model. We have also provided a first calculation of the long-range exponents $X, Y$, and $Z$, governing the decay of the critical amplitudes with the range of interactions. The calculation of these exponents for the two-dimensional $S$-state Potts model with long-range interactions will be of interest in order to provide a comparison with our results, and to verify whether the conjecture by Grinstein et al \cite{grinstein1985}, which states that reversible and irreversible models with the same symmetry are in the same universality class, can be extended to model systems with long-range interactions.

\begin{acknowledgments}
We thank the High Performance Computing Center at UFRN for providing the computational facilities to run the simulations. C.I.N. Sampaio Filho acknowledges financial support from the Brazilian agencies FUNCAP and CAPES. 
\end{acknowledgments}

\end{document}